\documentclass[
twocolumn, amsmath, amssymb, floatfix,
nofootinbib,wrapfloat byrevtex]{revtex4}

\usepackage{amsmath,mathrsfs,graphicx,epstopdf,placeins,float}
\usepackage{booktabs}
\usepackage[unicode]{hyperref}
\usepackage{multirow}
\usepackage{pstricks}

\newcommand{\beq}[1]{\begin{equation}\label{#1}}
\newcommand{\eeq}{\end{equation}}
\newcommand{\bea}[1]{\begin{eqnarray}\label{#1}}
\newcommand{\eea}{\end{eqnarray}}

\begin{document} 
	
\title{Chaotic D1-D5 Black Hole Dynamics through Networks}
\author{Han-qing Shi}
\email{hq-shi@emails.bjut.edu.cn}
\author{Xiao-yue Sun}
\email{xy$_$sun@emails.bjut.edu.cn}
\author{Ding-fang Zeng\footnote{corresponding author}}
\email{dfzeng@bjut.edu.cn}
\affiliation{Theoretical Physics Division, College of Applied Sciences, Beijing University of Technology}
	
\begin{abstract}
This work studies dynamics controlling the transition between different microstates of two charge D1-D5 black holes by network methods, in which microstates of the system are defined as network nodes, while transitions between them are defined as edges. It is found that the eigenspectrum of this network's Laplacian matrix, which is identified with Hamiltonians of the microstate system, has completely the same Nearest-Neighbor Spacing Distribution as that of general Gaussian Orthogonal Ensemble of Random Matrices. According to the BGS, i.e. Bohigas, Giannoni and Schmit conjecture, this forms evidence for chaotic features of the D1-D5 microstate dynamics. This evidence is further strengthened by observations that inverse of the first/minimal nonzero eigenvalue of the Laplacian matrix is proportional to logarithms of the microstate number of the system. By Sekino and Susskind, this means that dynamics of the D1-D5 black hole microstates are not only chaotic, but also the fastest scrambler in nature.
\end{abstract}
	
\maketitle

{\bf\em Motivation and conclusion} Understanding dynamics of the black hole microstate is a key step towards the final theory of quantum gravitation. Many important progresses on this direction have been achieved in string theory and Anti deSitter space/Conformal Field Theory \cite{StromingerVafa,AdSCFT,Mathur1,Mathur2,BW,LM,LLM} correspondences, AdS/CFT hereafter, loop quantum gravity \cite{Rovelli,Ashtekar}, as well as other approaches such as the semiclassic idea of \cite{Zeng1,Zeng2,Zeng3,Zeng4}.  For large mass black holes, the number of microstate available to the system is huge and considerable complicating relations must occur among them. Network method \cite{Networks} is a natural tool to explore systems like this which involve many degrees of freedoms and complicated relations. A. Charles and D. Mayerson \cite{BHnet} considered random walks on the microstate network of black holes to study their evolution and see interesting late-time behaviors of the system. 

Since the work of Hayden and Preskill \cite{HP}, relations between the black hole microstate dynamics and its information processing efficiency and chaotic feature arouse wide interests on the market. Sekino and Susskind \cite{scramblebh} suggest that black holes may be the fastest scrambler in nature.  Maldacena, Shenker and Stanford show that black holes are maximally chaotic in the sense that a bound on the early time behavior of OTOC is saturated \cite{MSS}. Investigations of  \cite{SYK,Kitaev,MS2016} uncover many similarities between the Sachdev-Ye-Kitaev model and black holes and provide more precise and quantitative supports for chaotic features of the black hole dynamics. Basing on this model and through analytical continued partition function as well as correlation functions, Cotler et. al \cite{BHRM,CCRT} numerically established that the late-time behavior of horizon fluctuations in large AdS black holes is governed by random matrix dynamics characteristic of quantum chaotic systems. 

Our interest in this work is chaotic features of the D1-D5 black hole dynamics \cite{Mathur1,Mathur2,D1D5review}. The CFT side and weak coupling limit of this issue are discussed in \cite{BCCS1703}, we will focus on its AdS side and general coupling strength aspect, but in string gas approximation. The network idea \cite{BHnet,CTQW,FG,CTRW} allows us to extract nontrivial information about the dynamic without knowing details of the interaction. In this idea the D1-D5 microstates are taken as nodes and transitions among them as edges of the network. We discard the late time behavior of any initial state's evolution as the investigating goal \cite{BHnet}, but take the network Laplacian matrix as the key object and identify it as the hamiltonian controlling transitions among different microstates. We observe that
\begin{enumerate}
\item[(i)] the Nearest-Neighbor Spacing Distribution(NNSD) of the Laplacian matrix's eigenvalue, follows universal statistics of the Gaussian Orthogonal Ensemble(GOE) of Random Matrices;
\item[(ii)] the system's typical relaxation time or inverse of the minimal nonzero eigenvalue of the Laplacian matrix manifests linear feature with logrithms of the system's microstate number.
\end{enumerate}

According to the BGS conjecture which states that systems whose NNSD of hamiltonian spectrum exhibits equal statistics as the GOE matrices will be chaotic in the classic limit, our observations (i) forms direct bulk space evidence for chaotic features of the D1-D5 black hole dynamics in the strong coupling limits, while (ii) provide strong supports for the fastest scrambling ability of black holes in the nature. 
Relations between the information processing ability of black holes and community structures of their microstate network are explored before summarizing section. Existence of communities in a network implies that microstates in one subnet are connected more tightly than those are connected among different sub-ones.

\begin{figure}[h]
\begin{center}
\includegraphics[scale=0.41]{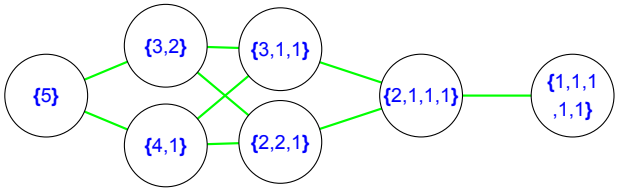}
\end{center}
\caption{The network representation of all microstates of D1-D5 black holes with $N_1\cdot N_5=5$}
\label{network5pt}
\end{figure}

{\bf\em Network definition and microstate dynamics} The D1-D5 system \cite{D1D5review,BW} is the most well understood string theory model for black hole microstates. Constraining by computation resources, we focus on in the current work the two charge system for simplicity. The same research to three charge systems involves no principle difficulties. The system lives in a $R^{4,1}{\times}S^1{\times}T^4$ spacetime, with the $N_1$ D1-brane wraps on $S^1$ while the $N_5$ D5-brane wraps on the $S^1{\times}T^4$. In the so called ``string gas" picture, the whole system can be considered as variable $\{N_w\}$ strands of short string, each wound the $S^1$ dimension $\{w\}$ times, with the total winding number fixed as
\begin{equation}
\sum_{w=1}^{\infty}wN_{w}=N_1{\cdot}N_5\equiv N
\end{equation} 
This allows us to represent each microstate of the system with an unordered set $\{w_{1},w_{2},\ldots\}$, each consisting of an integer partition of $N$ and forming a node in the microstate network. These microstate could be transformed into each other through a parent string's splitting or children string's merging. This will be represented with edges on the network. FIG.\ref{network5pt} gives an example of network representation for a black hole of 5 possible microstate. It is very natural to understand that sizes of the network will grow rapidly with $N$. 

Obviously, dynamics of the string interaction will affect the structure of the network directly, e.g. the linking strength of different edges. However, it is found \cite{BHnet} that although the adjacency matrices $A_{ij}=\omega(i) \Gamma(i{\rightarrow}j) \omega(j)$ affect the network structure heavily,  where $\omega(i)$ is the string state degeneracy and $\Gamma(i{\rightarrow}j)$ is the transition rate between $i$-$j$ state. The concrete form of $\omega(i)$ and $\Gamma(i{\rightarrow}j)$ or details of the string interaction dynamics on the late time behavior of general initial configurations of the network are unimportant. In other words, it is the inter-nodes being linked or not, rather than their linking strength's being robust or fragile that determines behavior of the network.  Basing on this fact, we will in this work simply set the network adjacency matrices $A_{ij}$ to be 1 or 0, depending on nodes $i$-$j$ could be connected or not through one event of a parent string's splitting or two children string's merging, with the winding number conservation as the only judging criteria.  Obviously this value assignment scheme for $A_{ij}$ is independent of the string coupling strength. So physics extracting from their spectrums is valid in general coupling strengths. Of course, string gas approximation is still a qualification on our works' value.

According to \cite{CTQW,FG}, evolutions of quantum state in the Hilbert space can be written as diffusions on the network controlled by matrix $A_{ij}$. Starting with an arbitrary initial state $\sum_ip_i|i{\rangle\langle}i|$, the system will evolve the following way
\begin{equation}
\dfrac{dp_{i}}{dt}=C\sum_{j}A_{ij}(p_{j}-p_{i})=C\sum_{j}(A_{ij}-\delta_{ij}d_{i})p_{j},
\end{equation}
where $C$ is the diffusion constant and $d_{i}\equiv\sum_{j}A_{ij}$ is the node degree. Writing this in the standard form of diffusion or transportation equations on the network \cite{Networks}, we have, where ${\bf L}$ is named the Laplacian matrix
\begin{equation}
\dfrac{dp}{dt}+C{\bf L}p=0
\label{eqlaplace},
\end{equation}
obviously ${\bf L}=D-A$,  with $D$ being the diagonal node degree matrix and $A$, the adjacency matrix. Denoting the eigenstates of ${\bf L}$ as $\left|q_{n}\right\rangle$, the solutions of \eqref{eqlaplace} can be formally written as, where $|\mathrm{ini}\rangle$ represent an arbitrary initial state,
\begin{equation}
p_{i}(t)={\langle}i|e^{-C {\bf L} t}|\mathrm{ini}\rangle
=\sum_{n} e^{-\lambda_{n} C t}{\langle}i|q_{n}\rangle\langle q_{n}|\mathrm{ini}\rangle.
\label{piSol}
\end{equation}
So the eigenvalue and eigenvectors of the Laplacian matrix determine evolutions of the diffusion process. It can be proven that ${\bf L}$ is non-negative and has one zero eigenvalue at least \cite{Networks}. Denote the spectrum in ascending order $\lambda_{0}\leq\lambda_{1}\leq\cdots$, as $t\rightarrow\infty$, the only term surviving in \eqref{piSol} is the $\lambda_{0}=0$ term. This means all diffusions on the network has a final equilibrium. While inverse of the first nonzero eigenvalue $\lambda_1^{-1}$  characterizes the relaxation time of the system. 

All microstates represented by nodes in our network are of equal mass/energy. They are eigenstates of free string hamiltonian. However, the hamiltonian identified with the network Laplacian matrix contains interaction. It is this interaction that makes the string-string merging, or one to two string splitting process possible, i.e.
\begin{equation}
C{\bf L}\equiv{\bf H} ={\bf H}_{0}+{\bf H}_{int}.
\end{equation}
${\bf H}_{0}$ here is the Hamiltonian whose eigenstates consist the network nodes, while ${\bf H}_{int}$ is the interaction which makes the transition between different microstates possible
\begin{equation}
{\bf H}_{0}\psi_{i}=E\psi_{i},~
{\bf H}\psi_{i}=C{\bf L}_{ij}\psi_{j}.
\end{equation}
The string interaction follows from the non-extremality introduced of the otherwise exactly BPS supersymmetric D1-D5 brane system. Relative to ${\bf H}_0$, ${\bf H}_\mathrm{int}$ is not small quantities in the strong coupling limit. This will help to explain why spectrums of ${\bf H}$ exhibit wide spreading instead of narrow localizing profile in the following FIG.\ref{figrhol}.  But because ${\bf H}_\mathrm{int}$ takes effects only when two strings contacting each other, which happens very seldom in string gas approximations, we neglect its effects when classifying microstate of the system and characterize them by two-charges, i.e. the D1- and D5-charge, or NS1 winding number and Kaluza-Klein momentum number in dual theories. 

\begin{figure}[h]
\begin{center}
\includegraphics[scale=0.4]{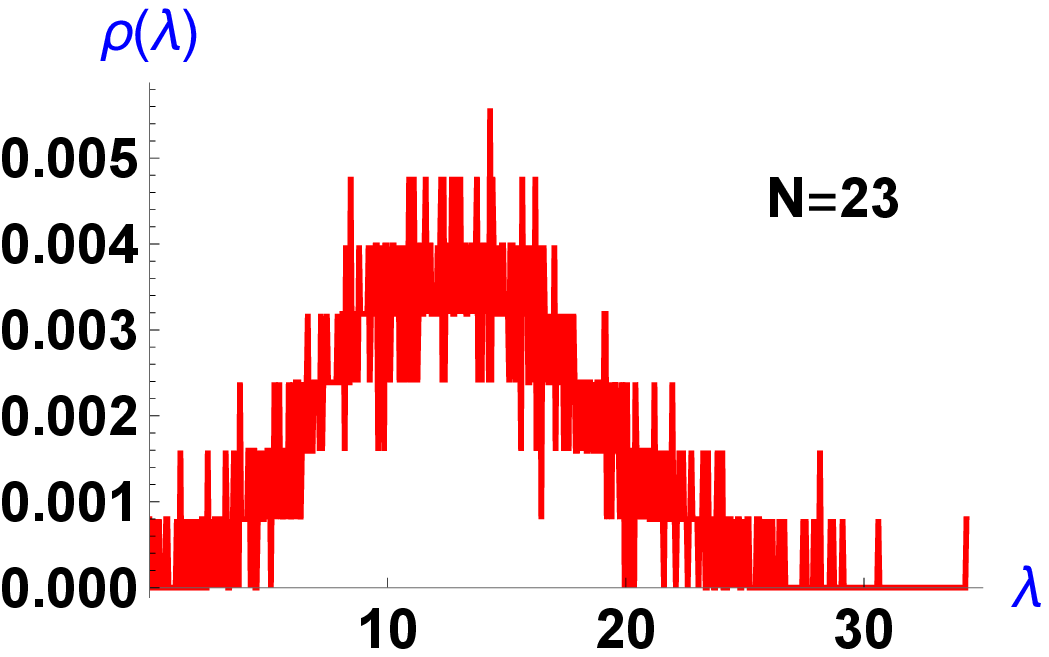}
\includegraphics[scale=0.4]{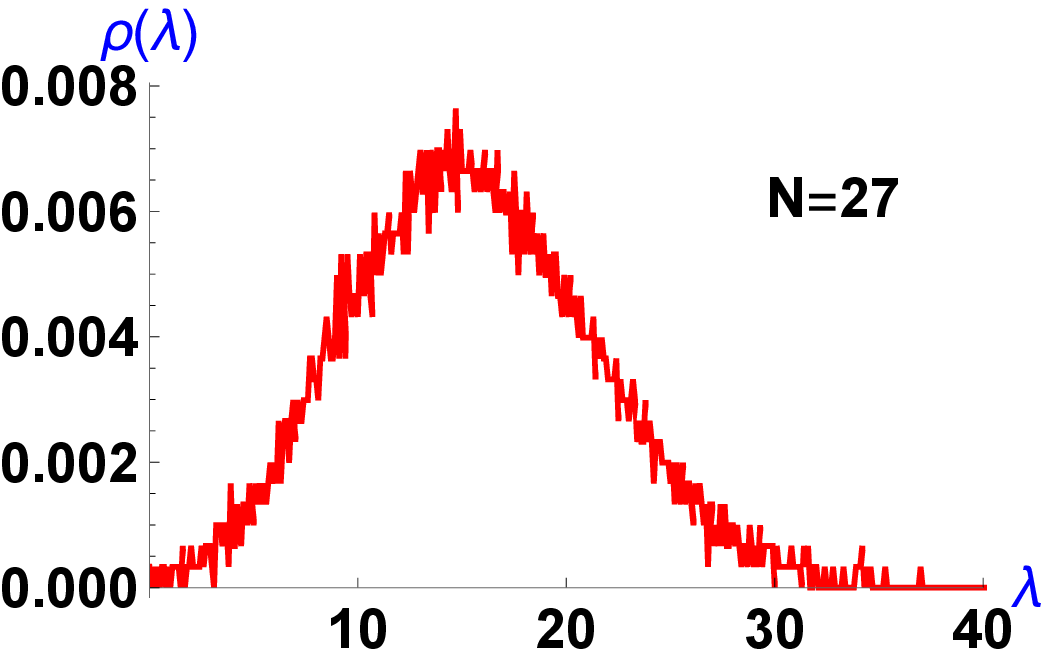}
\includegraphics[scale=0.4]{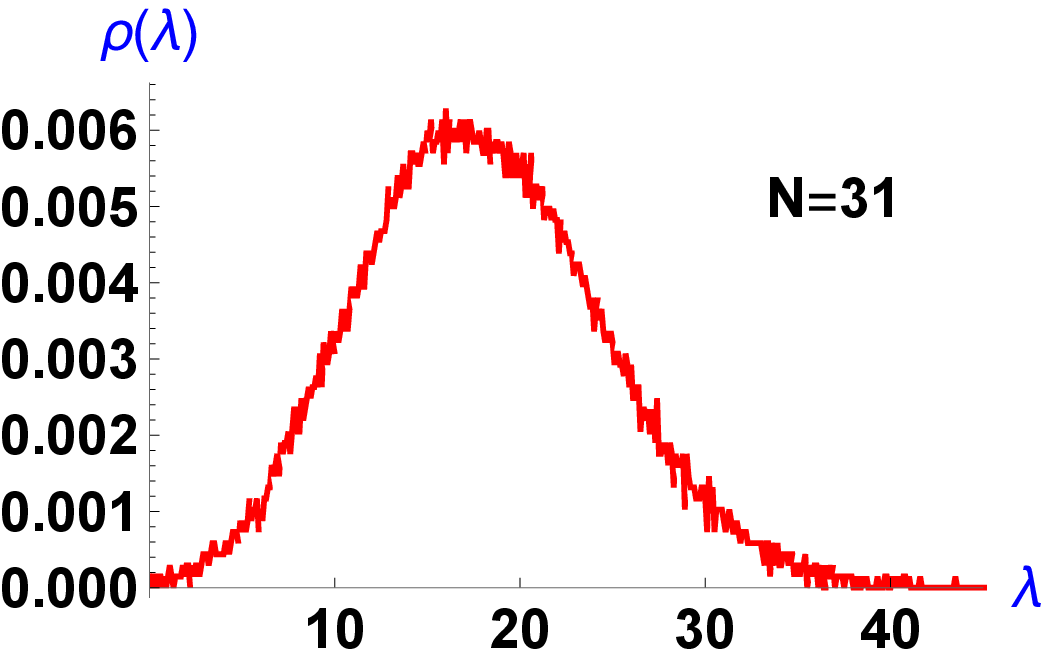}
\includegraphics[scale=0.4]{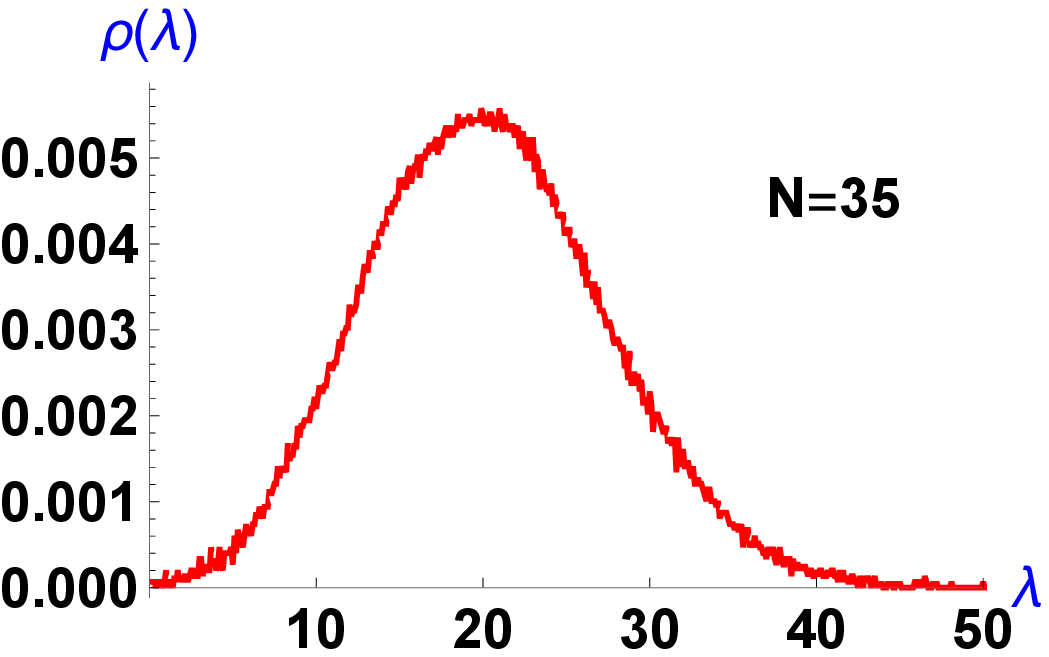}
\end{center}
\caption{The eigenvalue  distribution of Laplacian matrices of four D1-D5 black hole microstate network with $N=23,27,31$ and $35$. The horizontal axis is the possible eigenvalues and the vertical axis is their relative appearance frequency.}
\label{figrhol}
\end{figure}
\begin{figure}[h]
\begin{center}
\includegraphics[scale=0.4]{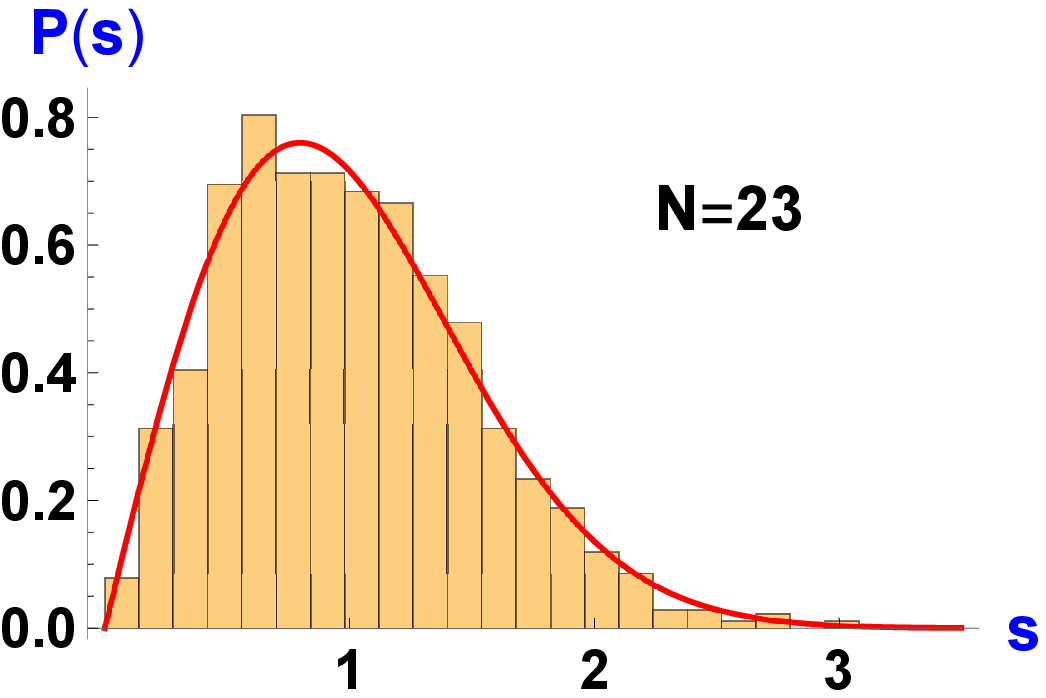}
\includegraphics[scale=0.4]{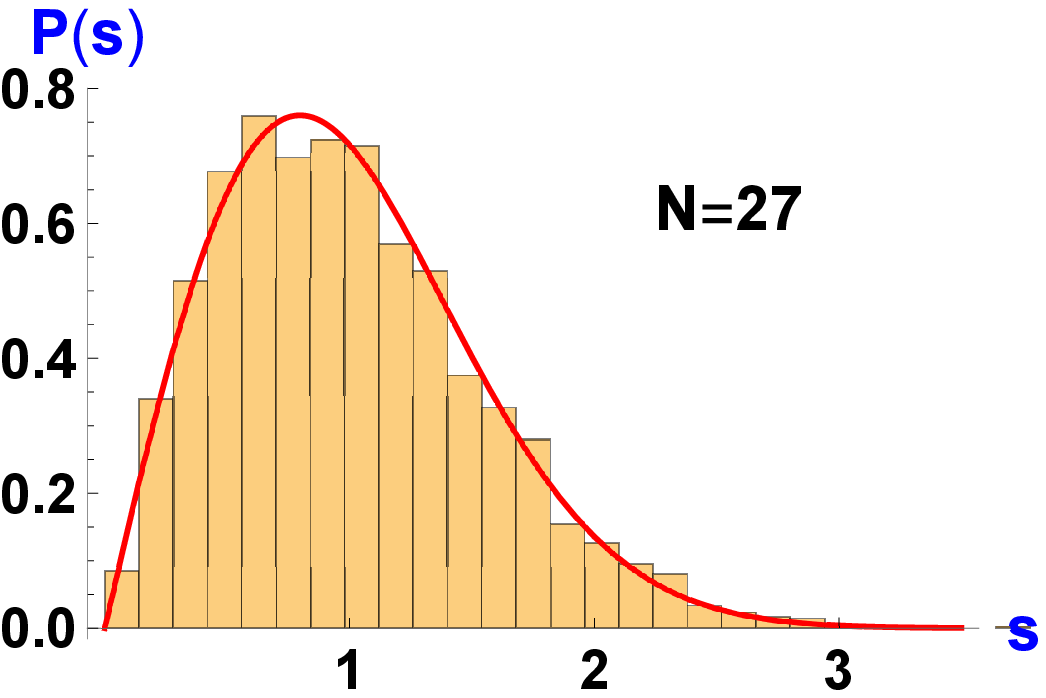}
\includegraphics[scale=0.4]{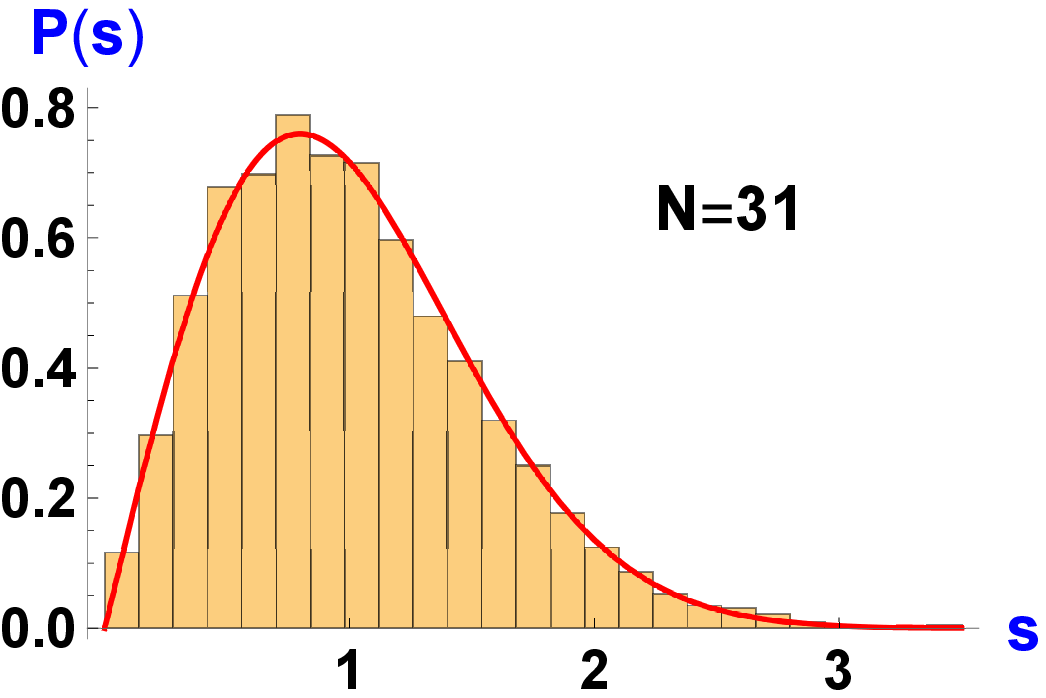}
\includegraphics[scale=0.4]{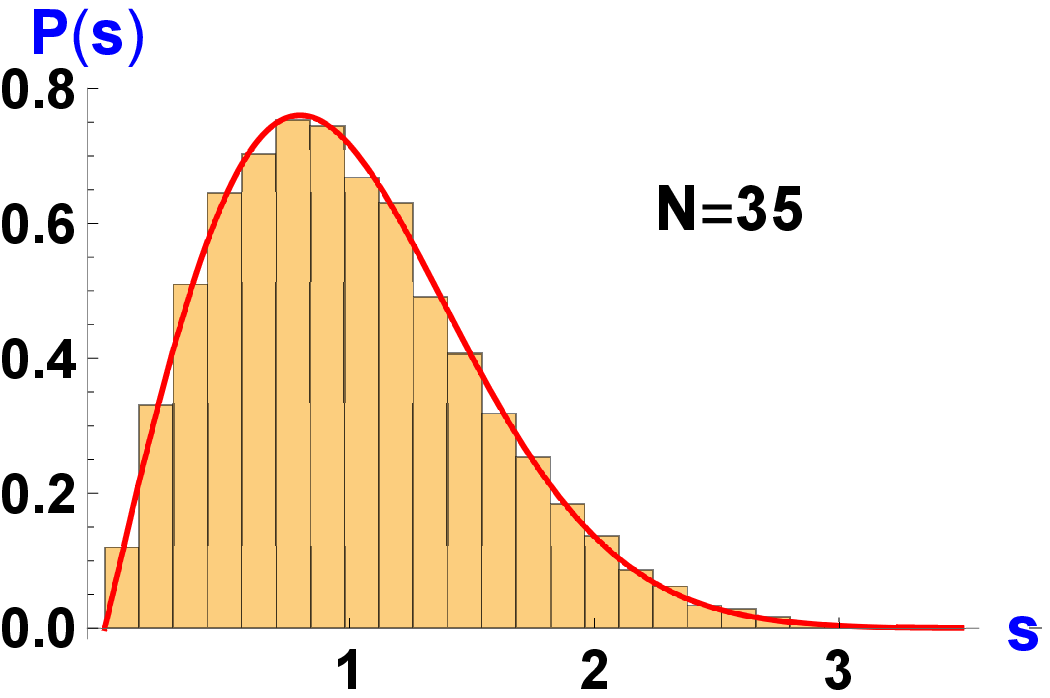}
\end{center}
\caption{The NNSD of eigenvalues of the Laplacian matrix of four typical D1-D5 black hole microstate network, the red curve is the distribution of eigenvalue spacing of GOE. The horizontal axis is the unfolded eigenvalue appearance frequency and the vertical axis is their relative distribution.}
\label{figPdl}
\end{figure}

{\bf\em Evidence for chaotic dynamics} We display in FIG.\ref{figrhol} and \ref{figPdl} the eigenvalue density/distribution $\rho(\lambda)$ of Laplacian matrices for four typical D1-D5 microstate network and their Nearest Neighbor Spacing Distribution $P(s)$ explicitly.  NNSD \cite{unfolding1,unfolding2}  is the distribution of relative appearance time/frequency $s$, $s$ itself has already the meaning of appearance time/frequency of $\lambda$, unfolded from the accumulated eigenvalue distribution 
\beq{}
\overline{n}(\lambda)=\int_{-\infty}^{\lambda} d \lambda^{\prime} \rho\left(\lambda^{\prime}\right)
,~
s_i=\overline{n}(\lambda_{i+1})-\overline{n}(\lambda_{i}).
\end{equation}
So the NNSD $P(s)$ is a kind of appearance frequency of appearance frequency. 

Very interestingly, our numeric results indicate that the NNSD of D1-D5 microstate network, matches exactly with those of the Gaussian Orthogonal Ensemble of Random Matrices., which is known as
\begin{equation}
P(s)=\frac{\pi}{2} s \mathrm{e}^{-\frac{\pi}{4} s^{2}},
\label{GOEps}
\end{equation}
By the so called BGS conjecture, for time-inverse invariant systems if their quantum hamiltonian has NNSD the same as those of GOE, they will manifest chaotic features in classic limit. This means that dynamics controlling the transition among different microstates of our D1-D5 brane system is chaotic in nature. This is a highly nontrivial but exciting results, because our network building process introduces nowhere random ingredients directly. Obviously, our results forms direct bulk space evidences for chaotic features of the black hole dynamics, while other works \cite{MS2016,BHRM,CCRT,BCCS1703} accomplish this goal almost exclusively from the dual gauge theory side.

\begin{figure}[h]
\begin{center}
\includegraphics[scale=0.65]{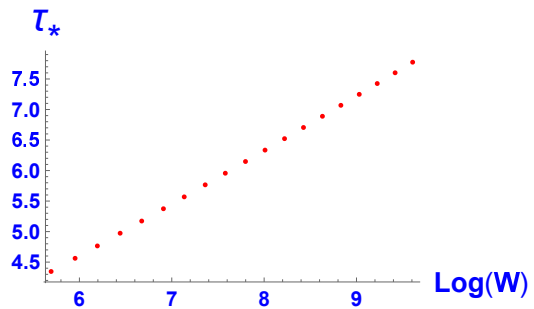}
\end{center}
\caption{Relations between the inverse of first nonzero eigenvalue $\lambda_{1}^{-1}$ of network Laplacian matrix versus logarithms of the number of microstates $\log(W)$ of the D1-D5 two charge black holes.}
\label{figtaus}
\end{figure}
Chaotic features of black hole dynamics are also related with its ability of information processing. By reference \cite{scramblebh} black hole is the fastest scrambler in the nature, which means that information fallen into them will be scrambled uniformly inside in times logarithmically proportional with the system's number of degrees of freedom. In our two charge D1-D5 microstate networks, the network Laplacian matrix is identified with the Hamiltonian controlling the transition among different states. So information scrambling time $\tau_\star$ of the system must be corresponded with the inverse of first, i.e. minimal nonzero eigenvalue $\lambda_1^{-1}$ of the Laplacian matrix. We display in FIG.\ref{figtaus} variations of $\tau_\star$ with respect to the logarithmic value of the corresponding microstate number for several black holes, with $N_1\cdot N_5$ ranges from 17 to 35. Obvioustly, our results forms strong support for the fast scrambler conjecture of reference \cite{scramblebh}.

It is worth emphasizing that the two key observations extracted from the Laplacian matrix spectrum in this section are valid in general string coupling conditions. Although string gas approximation represents a new constraint on our observations' validity, it is a more loose one than weak coupling limits.

\begin{figure}[h]
\begin{center}
\includegraphics[scale=0.5]{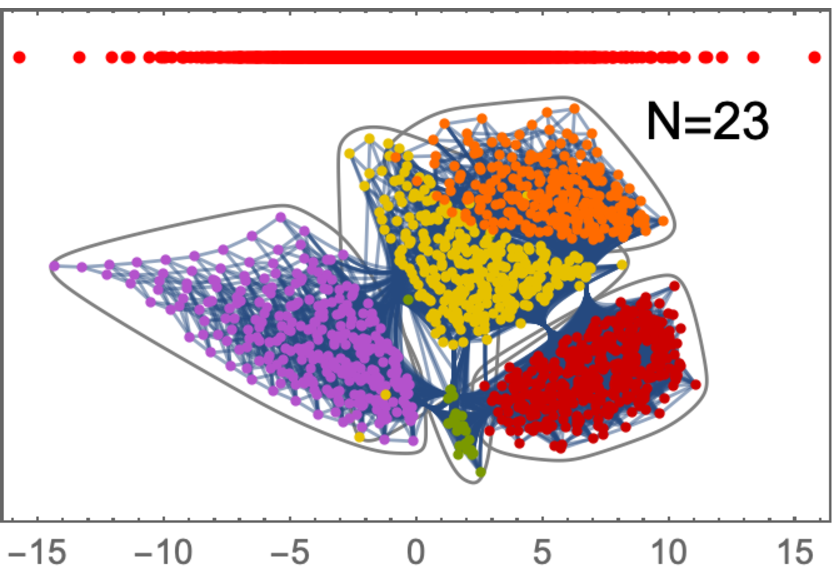}
\includegraphics[scale=0.5]{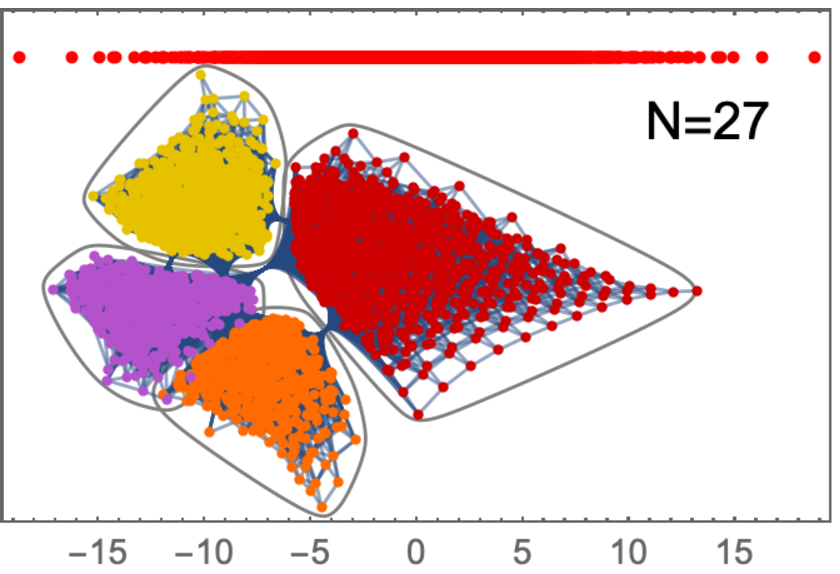}
\includegraphics[scale=0.5]{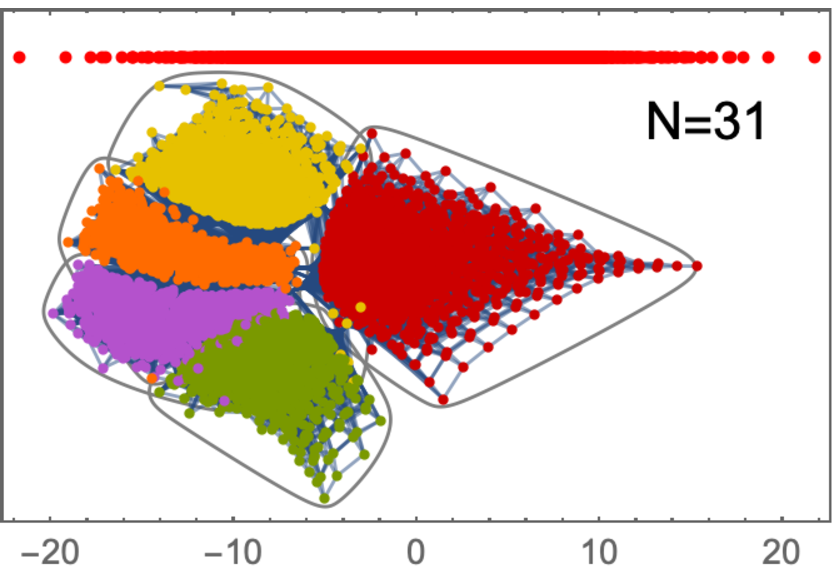}
\includegraphics[scale=0.5]{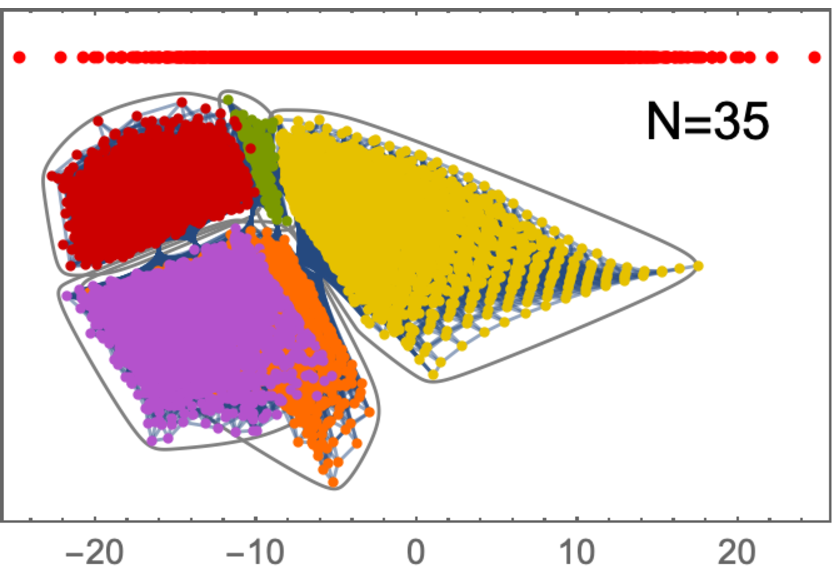}
\end{center}
\caption{Communities of microstate network of four different size D1-D5 black holes $N=23, 27, 31, 35$ and the adjacency matrix eigenvalue's aggregation feature, the upper horizontal scattering points in red.}
\label{figcomm}
\end{figure}

{\bf\em Community and information processing efficiency} Community \cite{Networks,Communitydetection} or subnetwork structure is another interesting feature of general networks. However, exactly defining and identifying community in a general network is a very difficult question mathematically. In practice, studying spectrums of the network adjacency matrix $A_{ij}$ may be the most convenient way \cite{Scd}. We display in FIG. \ref{figcomm} the community structure and corresponding $A_{ij}$ eigenvalue distribution for 4 typical networks representing the microstate of D1-D5 black holes with $N\equiv N_1\cdot N_5=23,27,31,35$. From the figure we easily see that as $N$ increases the number of communities in the network equals more and more precisely the number of $A_{ij}$'s eigenvalues well separated from the big bulk of other eigenvalues, with the bulk itself also considered as a ``well-separated'' one. This coincides with the observation of \cite{Scd} remarkably well. 

\begin{figure}[h]
\begin{center}
\includegraphics[scale=0.33]{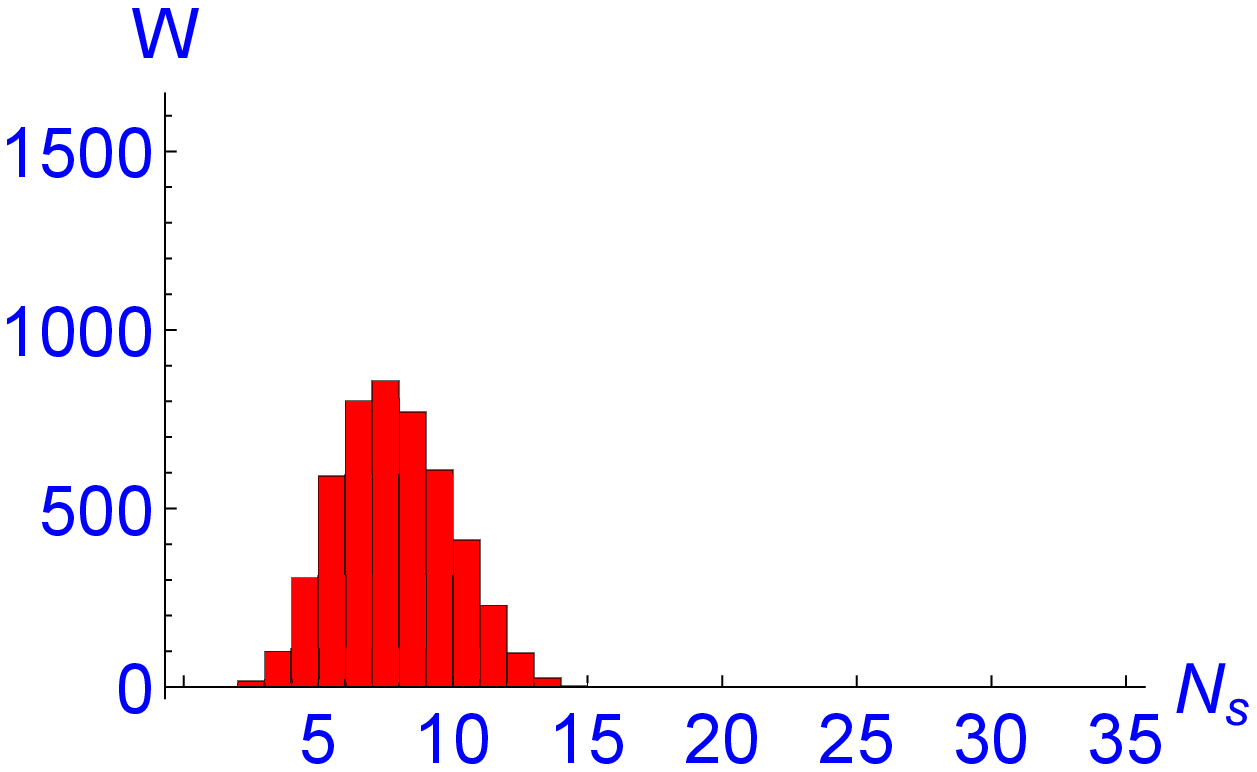}
\includegraphics[scale=0.33]{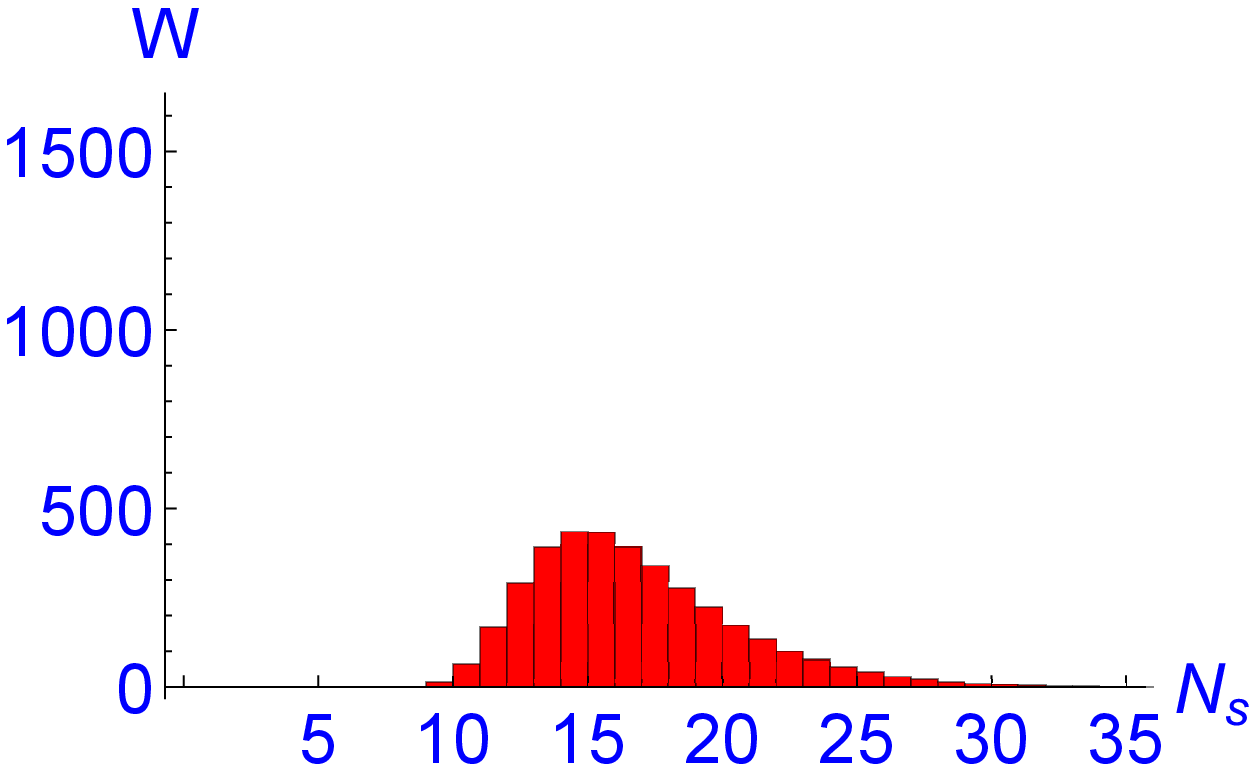}
\includegraphics[scale=0.33]{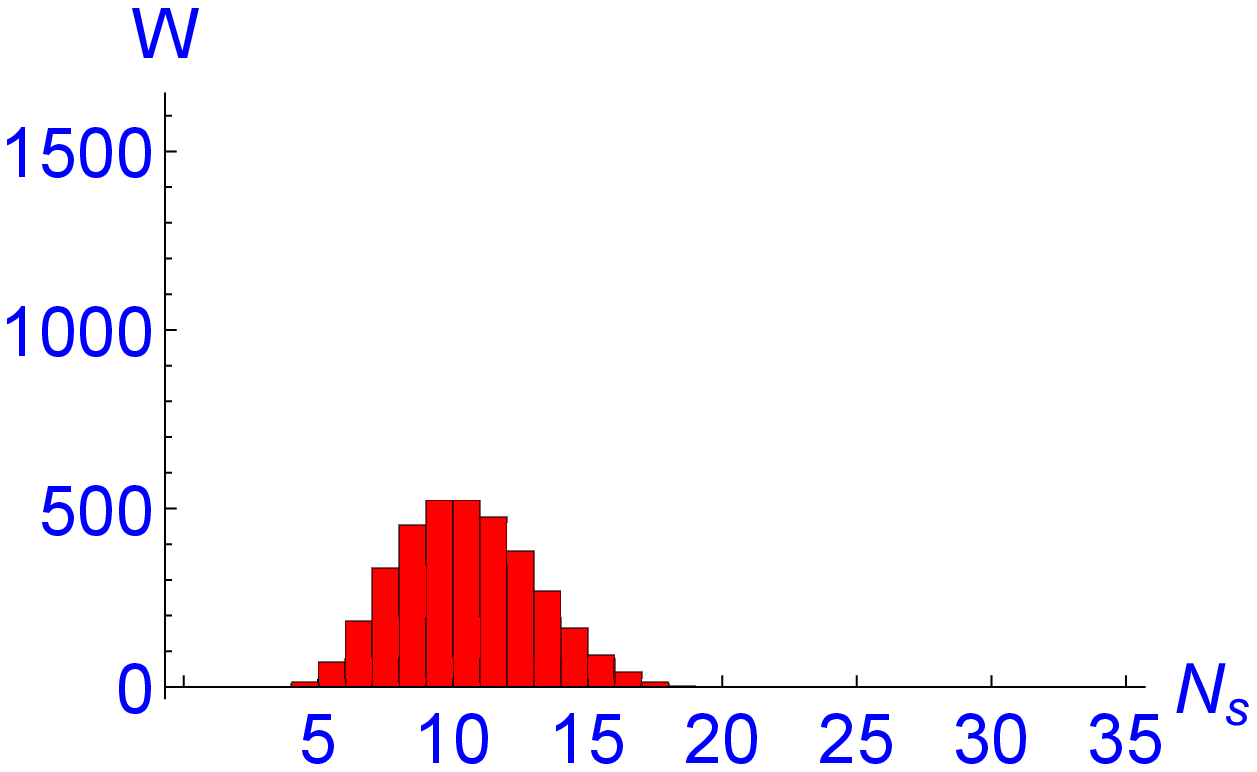}
\includegraphics[scale=0.33]{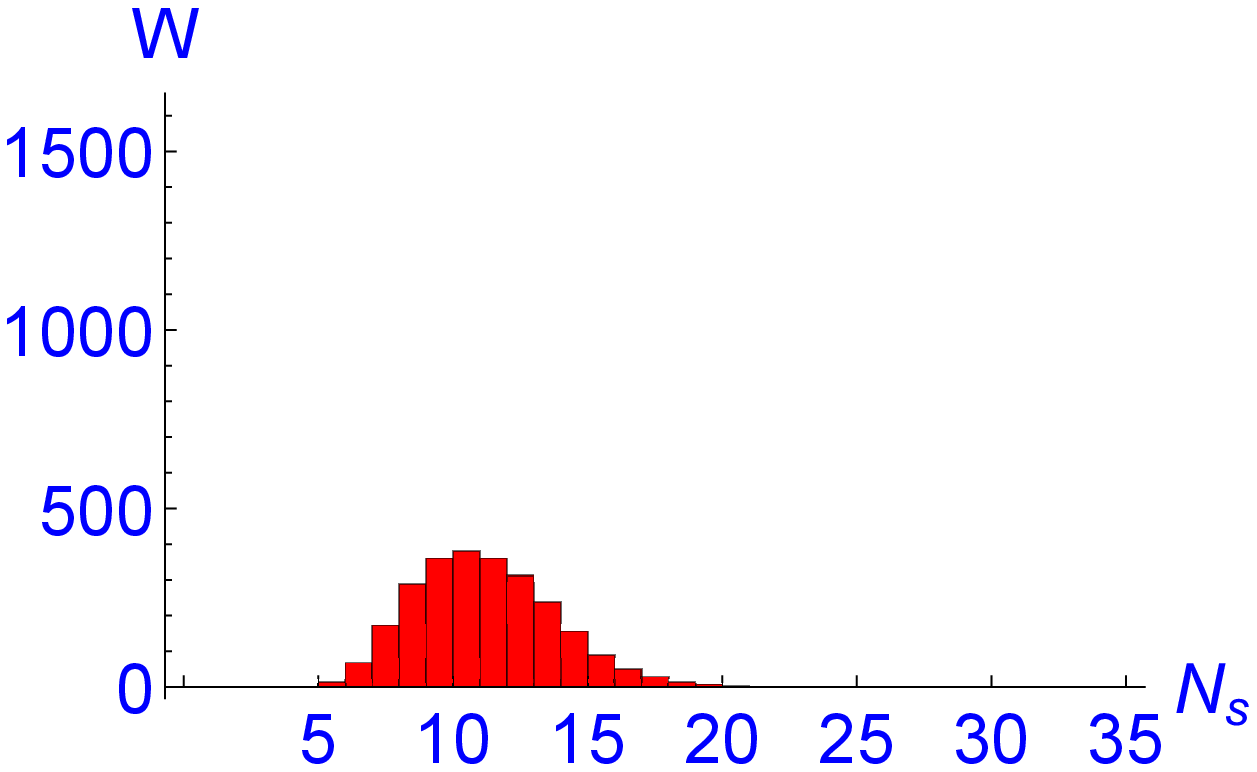}
\includegraphics[scale=0.33]{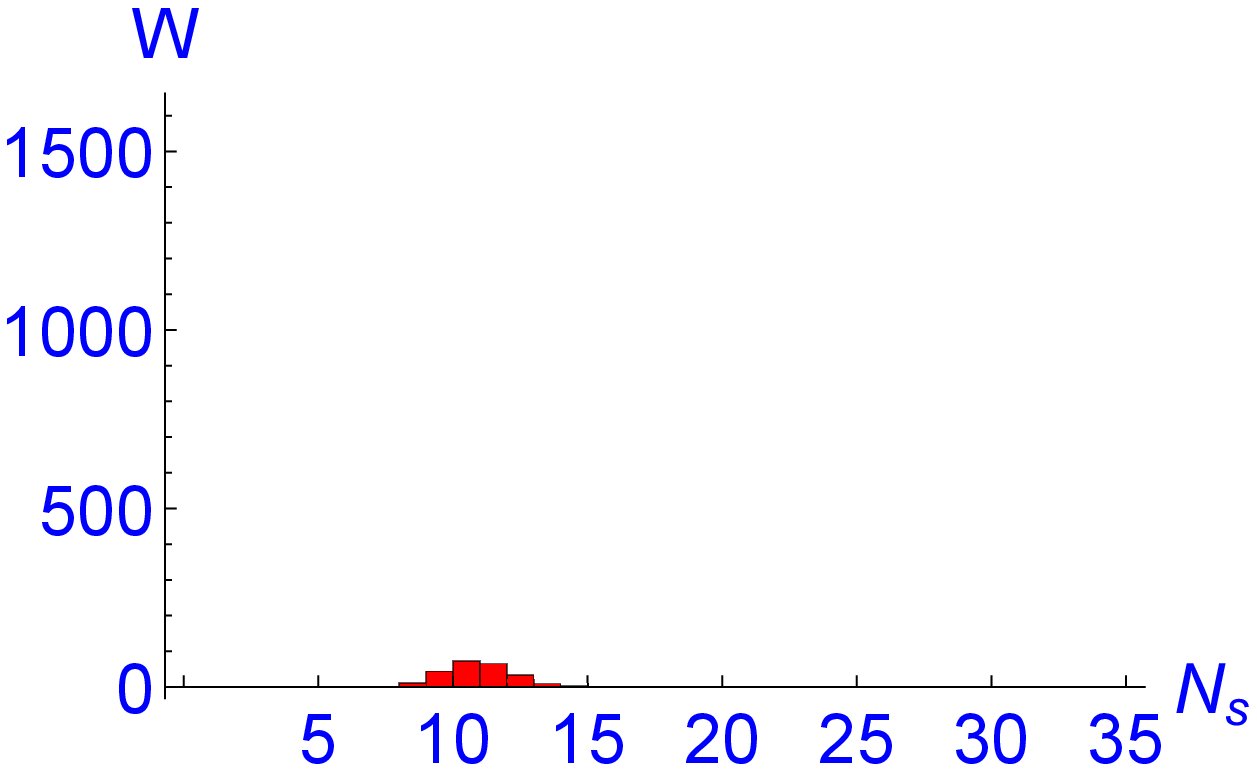}
\includegraphics[scale=0.33]{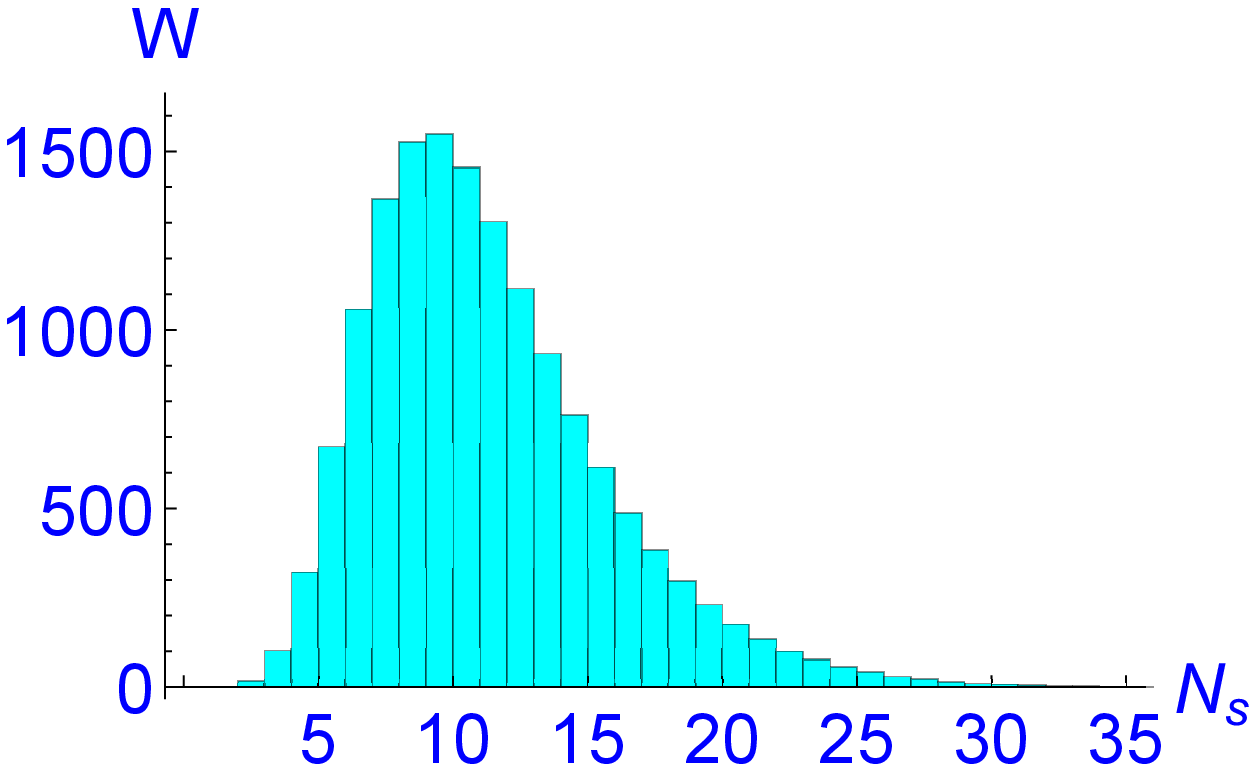}
\end{center}
\caption{The distribution of the number of short strings in the five community or subnetworks consisting the D1-D5 microstate network and the whole network itself, the last one, with total mass/energy parameter $N=35$. The horizontal axis is the number of short strings $N_{s}$ while the vertical one is the number of microstate $W$ with $N_s$ short strings.}
\label{figNscomm}
\end{figure}

Intuitively, when communities appear in the D1-D5 network,  microstate in one community will be more frequently connected with others in the same community than with states in others. Basing on this intuition, it is reasonable to expect that community structure should have effects on information processing efficiency of the system.  Recalling that in the two charge D1-D5 system, microstates are defined essentially the way a long string's decomposition into many short ones. We display in FIG.\ref{figNscomm} the distribution of short string numbers in the five communities consisting the whole microstate network and the whole network itself with $N=35$. From the figure, we easily see that $W(N_{s})$ has always one peak only in all communities. This means that microstates in the same community have approximately the same number of short strings $N_{s}$. This is because such microstates can be more easily transformed into each other, are thus more tightly connected with each other. Obviously, information denoted by them will be more efficiently scrambled among each other. In comparison, different states separated by larger gap of short string numbers need longer sequence of transition to evolve into each other, so connections among them are less tightly.
 
\begin{table}[h]  
\begin{tabular}{|c|c|c|c|c|c|c|c|c|c|c|c|c|c|}  
\hline  
$N$ & 10 & 11 & 12 & 13 & 14 & 15 & 16 & 17 & 18 & 19 & 20 & 21 & 22 \\ 
\hline  
n.c. & 4 & 4 & 4 & 4 & 5 & 4 & 5 & 4 & 5 & 4 & 5 & 5 & 5 \\  
\hline  
$N$ & 23 & 24 & 25 & 26 & 27 & 28 & 29 & 30 & 31 & 32 & 33 & 34 & 35 \\ 
\hline  
n.c. & 5 & 5 & 4 & 5 & 4 & 5 & 5 & 5 & 5 & 5 & 5 & 5 & 5 \\  
\hline  
\end{tabular} 
\caption{Number of communities (n.c.) in the D1-D5 microstate network with the black hole mass/energy parameter $N$ ranges from 10 to 35}  
\label{tblNetwork}
\end{table}
In practical operations, division and identification of communities in general networks contain big uncertainties. We display in TABLE \ref{tblNetwork} the variation of community numbers in our two charge D1-D5 black hole microstate network as the total energy of the system increases. From the table, we easily see that  the number of communities in different size black holes approaches more and more closely to the uniform value 5, although fluctuation/errors appear unavoidably due to exact division and identification schemes' lacking. Since in our networks, positive and negative eigenvalues of $A_{ij}$ always appear symmetrically, the asymptotic value of the number should be 5 instead of 4. Since black holes are conjectured the fastest scrambler in nature, this maybe the smallest number of communities for networks representing general multi-degree dynamic systems.

{\bf\em Summary and discussion}
We construct simple network representation for dynamics of two charge D1-D5 black hole microstates. In our representation, elements of the network adjacency matrix are assigned values either 1 or 0, depending on related microstate nodes can be transformed into each other or not under the only constraint of energy conservation. We identify Hamiltonians controlling dynamics of the D1-D5 microstate with the network Laplacian matrix, and find that NNSD of their spectrum has almost exactly the same distribution as those of GOE random matrices. According to the BGS conjecture, this forms a direct and concrete evidence for chaotic features of the D1-D5 microstate dynamics. As further evidence, we find that inverse of the first/minimal nonzero eigenvalue of the network Laplacian matrix, when identified with the typical relaxation time of the system, manifest simple linear increasing law with the logarithm of the microstate number of the D1-D5 black holes. This means that dynamics of the D1-D5 microstate is not only chaotic, but also the fastest information scrambler in nature. We analyze relations between this high efficiency of information processing and the community structure of the network preliminarily. All our findings or analysis are made under string gas approximation but general coupling strengths of the underlying string theory.

Network method is a powerful tool for fundamental physics. For instance, in string landscape studies it provides dynamical mechanisms for vacuum selection \cite{Stringnet}.  As prospects for futures, firstly we suggest applying the method of this work to the more complex three charge D1-D5 black holes \cite{Mathur1,Mathur2} or black holes with nontrivial semi-classic inner structure and dynamics such as \cite{Zeng1,Zeng2,Zeng3,Zeng4}. Secondly, more aspects or characteristic quantities in  network theories corresponding to black hole microstate dynamics deserve investigations. For example, the network side concept of Mean-First-Passage-Time(MFPT) \cite{Noh} and complexity concepts \cite{Susskind1,Susskind2,Susskind3,Susskind4} on the black hole side, share much similarities intuitively. The latter is defined as the minimal number of simple operations needed for $|A\rangle$ evolving to $|B\rangle$. While the physical meaning of MFPT $\langle T_{i j}\rangle$ is the mean value of the least walk steps from node $i$ to $j$. As long as relations between MFPT and complexity can be established, in either three charge D1-D5 black holes or other ones, we will get more concrete and direct verification of the CV \cite{Susskind3} or CA \cite{Susskind4} conjecture. 

{\bf\em Acknowledgements} We thank Anthony M. Charles and Daniel R. Mayerson for help us reproducing the results of \cite{BHnet}, which is a key stimulating reference for our work. This work is supported by NSFC grant {no.}11875082.


\begin{thebibliography}{99}
\bibitem{StromingerVafa}
A. Strominger and C. Vafa,
``Microscopic Origin of the Bekenstein-Hawking Entropy'',
{\em Phys.Lett. B} {\bf 379} (1996) 99-104.
\href{https://arxiv.org/abs/hep-th/9601029}{hep-th/9601029}
	
\bibitem{AdSCFT}
Maldacena,
``The Large N limit of superconformal field theories and supergravity'',
{\em Adv.Theor.Math.Phys.} {\bf2} (1998) 231-252.
\href{https://arxiv.org/abs/hep-th/9711200}{hep-th/9711200}
	
\bibitem{Mathur1}
S. Mathur,
``The Fuzzball proposal for black holes: An Elementary review'',
{\em Fortsch.Phys.} {\bf 53} (2005) 793-827.
\href{https://arxiv.org/abs/hep-th/0502050}{hep-th/0502050}
	
\bibitem{Mathur2}
S. Mathur,
``The quantum structure of black holes'',
{\em Class.Quant.Grav.} {\bf 23} (2006) R115.
\href{https://arxiv.org/abs/hep-th/0510180}{hep-th/0510180}
	
\bibitem{BW}
I. Bena and N. Warner,
``Black holes, black rings and their microstates'',
{\em Lect.Notes Phys.} {\bf 755} (2008) 1-92.
\href{https://arxiv.org/abs/hep-th/0701216}{hep-th/0701216}
	
\bibitem{LM}
O. Lunin and S. Mathur,
``AdS/CFT duality and the black hole information paradox'',
{\em Nucl.Phys.B} {\bf 623} (2002) 342-394.
\href{https://arxiv.org/abs/hep-th/0109154}{hep-th/0109154}
	
\bibitem{LLM}
H. Lin, O. Lunin and J. Maldacena 
``Bubbling AdS space and 1/2 BPS geometries'',
{\em JHEP} {\bf 0410} (2004) 025.	\href{https://arxiv.org/abs/hep-th/0409174}{hep-th/0409174}
	
\bibitem{Rovelli}
C. Rovelli,
``Black hole entropy from loop quantum gravity'',
{\em Phys.Rev.Lett.} {\bf 77} (1996) 3288-3291.
\href{https://arxiv.org/abs/gr-qc/9603063}{gr-qc/9603063}
	
\bibitem{Ashtekar}
A. Ashtekar et.al.,
``Quantum geometry and black hole entropy'',
{\em Phys.Rev.Lett.} {\bf 80} (1998) 904-907.
\href{https://arxiv.org/abs/gr-qc/9710007}{gr-qc/9710007}
	
\bibitem{Zeng1}
Ding-fang Zeng,
``Resolving the Schwarzschild singularity in both classic and quantum gravity'',
{\em Nucl.Phys.B} {\bf 917} (2017) 178-192.
\href{https://arxiv.org/abs/1606.06178}{1606.06178}
	
\bibitem{Zeng2}
Ding-fang Zeng,
``Schwarzschild fuzzball and explicitly unitary Hawking radiation'',
{\em Nucl.Phys.B} {\bf 930} (2018) 533-544.
\href{https://arxiv.org/abs/1802.00675}{1802.00675}
	
\bibitem{Zeng3}
Ding-fang Zeng,
``Information Missing Puzzle, Where Is Hawking's Error?'',
Nucl.Phys. B941 (2019) 665-679
{\em Nucl.Phys.B} {\bf 941} (2019) 665-679.
\href{https://arxiv.org/abs/1804.06726}{1804.06726}
	
\bibitem{Zeng4}
Ding-fang Zeng,
``Exact Inner Metric and Microscopic State of $AdS_{3}$-Schwarzschld BHs'',
\href{https://arxiv.org/abs/1812.06777}{1812.06777}
	
\bibitem{Networks}
M. Newman,
``Networks: An Introduction'',
Oxford University Press, Oxford, United Kingdom, March, 2010.
	
\bibitem{BHnet}
A. M. Charles and D. R. Mayerson,
``Probing Black Hole Microstate Evolution with Networks and Random Walks'',
\href{https://arxiv.org/abs/1812.09328}{1812.09328}
	
\bibitem{HP}
P. Hayden and J. Preskill,
``Black holes as mirrors: quantum information in random subsystems''
{\em JHEP} {\bf0709} (2007) 120.
\href{https://arxiv.org/abs/0708.4025}{0708.4025}
	
\bibitem{scramblebh}
Y. Sekino and L. Susskind,
``Fast Scramblers'',
{\em JHEP} {\bf0801} (2008) 065.
\href{https://arxiv.org/abs/0808.2096v1}{0808.2096}
	
\bibitem{MSS}
J. Maldacena, S.H. Shenker and D..Stanford,
``A bound on chaos'',
{\em JHEP} {\bf1608} (2016) 106.
\href{https://arxiv.org/abs/1503.01409}{1503.01409}

\bibitem{SYK}
S. Sachdev \& J. Ye,
``Gapless spin-fluid ground state in a random quantum Heisenberg magnet'',
{\em Phys. Rev. Lett.} {\bf70} (1993) 3339.
\href{https://arxiv.org/abs/cond-mat/9212030}{cond-mat/9212030}
	
\bibitem{Kitaev}
A. Kitaev, 
``A simple model of quantum holography" Talks given at the KITP, 
Apr.7, 2015 and May 27, 2015.
	
\bibitem{MS2016}
J. Maldacena \& D. Stanford,
``Remarks on the Sachdev-Ye-Kitaev model'',
{\em Phys. Rev. D} {\bf94} (2016) 106002.
\href{https://arxiv.org/abs/1604.07818}{1604.07818}
	
\bibitem{BHRM}
J. S. Cotler et.al.,
``Black holes and random matrices''
{\em JHEP} {\bf1705} (2017) 118.
\href{https://arxiv.org/abs/1611.04650}{1611.04650}
	
\bibitem{CCRT}
J. Cotler et.al.,
``Chaos, Complexity, and Random Matrices''
{\em JHEP} {\bf1711} (2017) 048.
\href{https://https://arxiv.org/abs/1706.05400}{1706.05400}

\bibitem{D1D5review}
J. R. David, G. Mandal and S. R. Wadia,
``Microscopic formulation of black holes in string theory'',
{\em Phys.Rept.} {\bf 369} (2002) 549-686.
\href{https://arxiv.org/abs/hep-th/0203048}{hep-th/0203048}

\bibitem{BCCS1703}
V. Balasuramanian, B. Craps, B. Czech et. al,
``Echoes of chaos from string theory black holes'',
{\em JHEP} {\bf 1703} (2017) 154.
\href{http://arxiv.org/abs/arXiv:1612.04334}{1612.04334}
	
\bibitem{BGS}
O. Bohigas, M. J. Giannoni, and C. Schmit
``Characterization of chaotic quantum spectra and universality of level fluctuation laws'',
{\em Phys. Rev. Lett.} {\bf52} (1984) 1.

\bibitem{CTQW}
O. Mülken and A. Blumen,
``Continuous-time quantum walkswalks: Models for coherent transport on complex networks'',
{\em Phys.Rept.} {\bf 502} (2011) 37-87.
\href{https://arxiv.org/abs/1101.2572}{1101.2572}
		
\bibitem{FG}
E. Farhi and S. Gutmann, 
``Quantum computation and decision trees'',
{\em Phys. Rev. A} {\bf58} (1998) 915.
\href{https://arxiv.org/abs/quant-ph/9706062v2}{quant-ph/9706062}
	
\bibitem{CTRW}
N. van Kampen,
``Stochastic Processes in Physics and Chemistry'',
North-Holland, Amsterdam, 1990.
	
\bibitem{unfolding1}
J. M. G\'{o}mez,
``Misleading signatures of quantum chaos'',
{\em Phys. Rev. E} {\bf66} (2002) 036209.
\href{https://arxiv.org/abs/nlin/0112014}{nlin/0112014}	
	
\bibitem{unfolding2}
A. A. Abul-Magd and A. Y. Abul-Magd,
``Unfolding of the Spectrum for Chaotic and Mixed Systems'',
{\em Physica A} {\bf369} (2014) 185-194.
\href{https://arxiv.org/abs/1311.2419}{1311.2419}	

\bibitem{Communitydetection}
S. Fortunato and D. Hric,
``Community detection in networks: A user guide'',
{\em Phys.Rept.} {\bf 659} (2016) 1-44.
\href{https://arxiv.org/abs/1608.00163}{1608.00163}

\bibitem{Scd}
S. Chauhan, M. Girvan, and E. Ott,
``Spectral properties of networks with community structure''
{\em Phys. Rev. E} {\bf80} (2009) 056114.

\bibitem{Noh}
J. Noh and H. Rieger,
``Random Walks on Complex Networks'',
{\em Phys. Rev. Lett.} {\bf 92} (2004) 118701.
\href{https://arxiv.org/abs/cond-mat/0307719}{cond-mat:0307719}

\bibitem{Susskind1}
L. Susskind,
``Second law of quantum complexity'',
{\em Phys. Rev. D} {\bf 97} (2018) 086015.
\href{https://arxiv.org/abs/1701.01107}{1701.01107}

\bibitem{Susskind2}
L. Susskind,
``Three Lectures on Complexity and Black Holes'',
\href{https://arxiv.org/abs/1810.11563}{1810.11563}

\bibitem{Susskind3}
L. Susskind,
``Computational Complexity and Black Hole Horizons'',
{\em  Fortsch.Phys.} {\bf 64} (2016)  24-43.
\href{https://arxiv.org/abs/1403.5695}{1403.5695}
	
\bibitem{Susskind4}
L. Susskind,
``Complexity, action, and black holes'',
{\em Phys. Rev. D} {\bf 93} (2016) 086006.
\href{https://arxiv.org/abs/1512.04993} {1512.04993}

\bibitem{Stringnet}
J. Carifio et. al.,
``Vacuum Selection from Cosmology on Networks of String Geometries'',
{\em Phys. Rev. Lett.} {\bf121} (2018) 101602.
\href{https://arxiv.org/abs/1711.06685}{1711.06685}

\end{thebibliography}
\end{document}